\documentclass[journal]{IEEEtran}
\ifCLASSINFOpdf
\else
\fi
\usepackage{multirow}
\usepackage{flushend}
\usepackage[pdftex]{graphicx,color}
\usepackage[square,comma,sort&compress,numbers]{natbib}
\usepackage{amsmath}
\usepackage{amssymb}
\usepackage{algpseudocode,algorithm}
\newtheorem{lemma}{Lemma}
\DeclareMathSizes{10}{8}{7}{6}
\IEEEaftertitletext{\vspace{-2\baselineskip}}
\begin{document}
\title{Massive Multiple Access Based on Superposition Raptor Codes for M2M Communications}
\author{Mahyar~Shirvanimoghaddam,~\IEEEmembership{Member,~IEEE,}
                 Mischa~Dehler,~\IEEEmembership{Fellow,~IEEE,}
                Sarah~J.~Johnson,~\IEEEmembership{Member,~IEEE}

\thanks{The material in this paper was submitted in part to IEEE International Symposium on Information Theory (ISIT), 2016.

M. Shirvanimoghaddam and Sarah. J. Johnson are with School of Electrical Engineering and Computer Science, The University of Newcastle, NSW, Australia (e-mail: \{mahyar.shirvanimoghaddam; sarah.johnson\}@newcastle.edu.au).

M. Dohler is with King's College London, UK (email: mischa.dohler@kcl.au.uk).}}
\maketitle

\begin{abstract}
Machine-to-machine (M2M) wireless systems aim to provide ubiquitous connectivity between machine type communication (MTC) devices without any human intervention. Given the exponential growth of MTC traffic, it is of utmost importance to ensure that future wireless standards are capable of handling this traffic. In this paper, we focus on the design of a very efficient massive access strategy for highly dense cellular networks with M2M communications. Several MTC devices are allowed to simultaneously transmit at the same resource block by incorporating Raptor codes and superposition modulation. This significantly reduces the access delay and improves the achievable system throughput. A simple yet efficient random access strategy is proposed to only detect the selected preambles and the number of devices which have chosen them. No device identification is needed in the random access phase which significantly reduces the signalling overhead. The proposed scheme is analyzed and the maximum number of MTC devices that can be supported in a resource block is characterized as a function of the message length, number of available resources, and the number of preambles. Simulation results show that the proposed scheme can effectively support a massive number of M2M devices for a limited number of available resources, when the message size is small.
\end{abstract}

\IEEEpeerreviewmaketitle

\section{Introduction}
\IEEEPARstart{T}{he} ever increasing demand of industries to automate their real-time monitoring and control processes and the popularity of smart applications to improve our everyday life will exponentially increase machine-to-machine (M2M) system deployments in the near future \cite{ChallRA_LTEM2M}. M2M communications aim to enable trillions of multi-role devices, namely machine-type communication (MTC) devices, to communicate with each other and the underlying data transport infrastructure with little or no human interaction \cite{TUbiq,M2MMInternet}. M2M communications have potentially diverse applications across different industries, including healthcare, the smart city market, logistic, manufacturing, process automation, energy, and utilities \cite{MTCSurvey}. This makes M2M communications one of the fastest-growing technologies in the field of telecommunications. ​According to an updated market forecast from ABI Research, the number of devices will more than double from the current level, with 40.9 billion forecasted for 2020 \cite{ABI}. Furthermore, Gartner estimates that the Internet of Things (IoT) will include 26 billion units installed by 2020, and by that time, IoT product and service suppliers will generate incremental revenue exceeding \$300 billion in services \cite{Gartner}. This demonstrates the strong motivation for cellular wireless technology providers to participate in this market \cite{CoverageM2MLTE}. On the other hand, the ubiquitousness of cellular networks is a major incentive for M2M application developers to adopt cellular networks for their numerous applications \cite{HybridRAandDataM2M}.

The latest cellular communication standard developed by the third generation partnership project (3GPP) is Long-Term Evolution (LTE), which provides a flexible communication architecture to enable reliable communication at a lower cost per bit and to accommodate the continuous growth in wireless cellular demand \cite{CoverageM2MLTE}. However, LTE cellular networks, which are originally designed and engineered for human-to-human (H2H) communications, have been considered not suitable to handle the unique characteristics of M2M applications \cite{HybridRAandDataM2M}. 

Many M2M applications and device types share a set of key attributes that have to be considered in the design of future wireless networks. These include, sporadic transmission of small data bursts (only few kbs), massive number of devices, and low power consumption to extend battery life.  Moreover, M2M devices and applications have diverse quality-of-service (QoS) requirements and traffic patterns \cite{Mahyar_TWC}. For these reasons, leading standardization bodies, such as 3GPP, have commenced work on satisfying these and other constraints while not sacrificing current cellular system usage for human-based applications \cite{ChallM2MAccess}. For example, 3GPP has already specified the general requirements for MTC applications and identified issues and challenges related to them, and several network and device modifications have been considered in the future release of LTE, referred to as LTE-Advanced (LTE-A) \cite{3GPPLTEA}. However, a dramatic improvement in efficiency requires major changes to the air interface and core network \cite{MTCSurvey}.

The random access channel (RACH) of LTE and LTE-A has been identified as a key area in which an improvement for MTC traffic is necessary \cite{ChallRA_LTEM2M}. In fact, the connection-oriented communication in the current LTE standard can induce excessive signalling overhead in the case of transmitting small-sized data for M2M communications, especially when a large number of M2M devices attempt to access cellular networks at the same time \cite{HybridRAandDataM2M}. Moreover, many M2M devices stay out of  connection to save energy except for communicating with the network and transmitting a small amount of signalling data. Therefore, cellular networks should focus on how to deal with a massive number of connection requests to initiate the network connection before data transmission, rather than data traffic from numerous devices \cite{SpatialGroupRAM2M}.

In the current LTE standard, the uplink channel is divided into two sub-channels, namely the physical random access channel (PRACH) for preamble transmission and signalling overhead and physical uplink shared channel (PUSCH) for data transmission. Random access (RA) is the first step in establishing an air interface connection to access the cellular network, where multiple users/devices transmit random access preambles in PRACH. In M2M communications, due to a massive number of MTC devices, a preamble is usually selected by more than one device, which are then allocated with the same PUSCH for the data transmission by the base station (BS). This is the first problem in RA for M2M communications which is called preamble collision on PRACH and results in PUSCH wastage as the BS cannot decode any data packet due to the co-channel interference. The frequent preamble collisions in M2M communications also leads to network congestion, unexpected delays, packet loss, high energy consumption, high signaling overhead, and radio resource wastage \cite{M2MRA}. The second problem is that even if each MTC device selects a preamble without collision, there may not be enough resource blocks (RBs) to be allocated to the devices by the BS to the respective PUSCH \cite{RANewRAM2M}.

The focus of existing studies on RA for M2M communications is mostly limited to the first problem. In this vein and to reduce the access delay in M2M communications, several overload control mechanisms have been proposed, including \textit{dynamic allocation} \cite{EnhanceLTE,AutRA}, \textit{slotted access}, \textit{group-based} \cite{TUbiq,EEMacessM2M}, \textit{pull-based}, and \textit{access class barring} \cite{PriorACB,ClassBarr}. Moreover, the authors in \cite{SpatialGroupRAM2M} proposed a novel RA scheme, where a cell coverage is spatially partitioned into multiple group regions based on their delay and additional preambles can be provided by reducing the cyclic shift size in RA preambles.  A further improvement on \cite{SpatialGroupRAM2M} can be achieved for fixed location MTC devices based on fixed timing alignment and prediction of the possible occurrences of collisions \cite{NovelFixed}. A review of several RA overload control mechanisms can be found in \cite{M2MRA}. Although these approaches can reduce the access collisions to a certain degree, most of them still suffer from very high access delays in highly dense networks.

Unfortunately, only few studies have considered the second problem. More specifically, \cite{RANewRAM2M} proposed a new RA strategy by attaching the device ID to the preamble sent by the MTC device, enabling the BS to detect the collision in the the RA process. Moreover, the device's message is sent as part of the scheduled message in the RA phase of the LTE standard, which reduces the signalling overhead. A Hybrid RA and data transmission protocol has also been proposed in \cite{HybridRAandDataM2M}, where the available resources are dynamically allocated for the PRACH and PUSCH according to the periodic estimation of the number of active M2M devices by the BS. Although, the proposed approach can reduce the signalling overhead, which is suitable for M2M applications with small data sizes, it cannot solve the preamble collision problem when a massive number of devices are attempting to access the network.

In this paper, we propose a novel RA strategy for M2M communications which provides major improvements in terms of access delay and QoS, by shifting from conventional identification/authentication based RA strategies. The proposed strategy aims to 1) minimize the access delay by enabling the collided devices to transmit at the same data channel, consisting of several RBs, 2) minimize the signalling overhead by signaling once for each group of devices which have selected the same preamble, and 3) minimize the resource wastage due to efficient usage of available resources. The proposed scheme contains two phases, the RA phase and the data transmission phase. The devices do not need to be identified by the BS in the RA phase; instead, the device ID is sent along with its message in the data transmission phase and later is decoded by the BS. In the proposed scheme, collided devices are transmitting at the same data channel by using the same Raptor code \cite{Raptor}. More specifically, a single degree distribution is used for Raptor codes in all the devices, which significantly simplifies the system design as the code is not dependent on the number of devices or network condition. The BS will need to know only the number of active devices in a data channel to perform the decoding and device identification. In fact, the received signal at the BS can be realized as a superposition of coded symbols sent from the devices, which is then shown to be capacity approaching, when an appropriate successive interference cancellation (SIC) is used for the decoding. This is particularly suitable for M2M communications with strict power limitations, especially when the data size is very small and the number of devices is very large; thus, low rate Raptor codes in the low SNR regime can be effectively used for their data transmission. The maximum number of M2M devices which can transmit at the same resource block is then characterized as a function of the data size and the available bandwidth. The proposed scheme shows an excellent performance in highly dense M2M networks, which makes it an excellent choice for future wireless technologies.

The rest of the paper is organized as follows. Section II represent the system model. Section III represents the proposed random access scheme. An overview on Raptor codes and the proposed data transmission strategy is presented in Section IV. The rate performance analysis of the proposed scheme and the weight coefficient design are studied in Section V. In Section VI, we shed light on some of important practical issues of the proposed scheme. Section VII shows the simulation results, followed by some concluding remarks in Section VIII.
\section{System Model}
\begin{figure}[!t]
\centering
\includegraphics[scale=0.42]{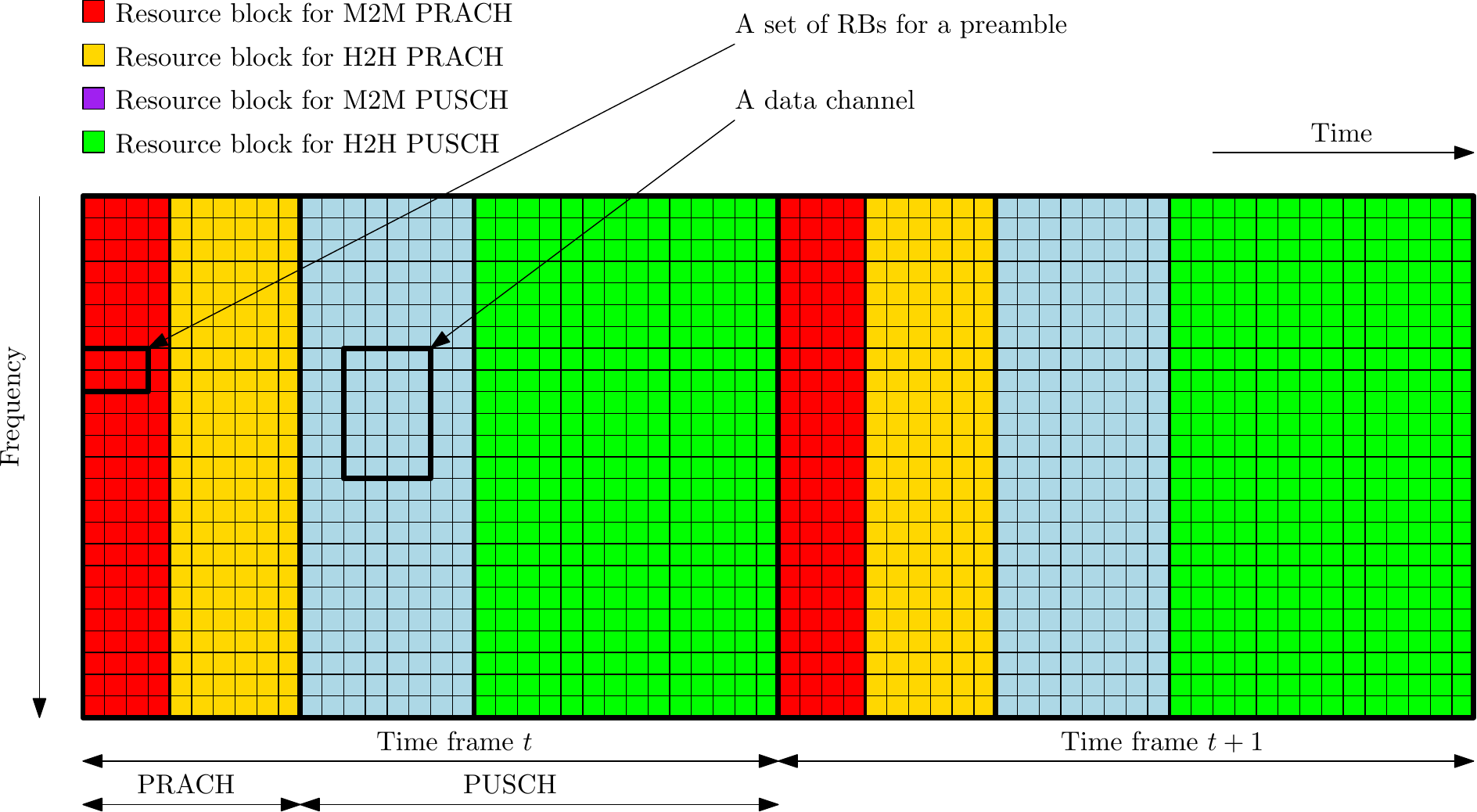}
\caption{Time-frequency domain model of uplink radio resource for the proposed scheme.}
\label{RBfiglabel}
\end{figure}

We consider a single-cell centered by a BS in a cellular wireless network, where M2M and H2H devices coexist and share radio resources. The number of M2M devices is assumed to be far greater than that of H2H devices. Similar to the LTE system, we consider the uplink of an orthogonal frequency division multiple access (OFDMA) system, where the radio resources are divided into units of resource blocks (RBs), each with time duration $\tau_\mathrm{s}$ and bandwidth $W_\mathrm{s}$. The time is divided into time frames of length $T_\mathrm{f}$, where the number of active M2M devices in each time frame is random and follows a Poisson process with rate $\lambda$ \cite{FunThroM2M}. Fig. \ref{RBfiglabel} shows the time-frequency model of the uplink radio resource for M2M communications. We also consider a contention-based random access strategy in M2M communications, where $N_\mathrm{s}$ different random access preambles are selected in a random access attempt \cite{M2MRA}.  The notation used in this paper is also summarized in Table \ref{NotSum} for quick reference.

We assume that the radio resources for M2M and H2H communications are separately managed. The radio resource manager can determine the number of required resources for M2M communications based on the information on the traffic loads of M2M and H2H communications. Traffic load information of H2H communications can be obtained in LTE based on the buffer status report and for M2M communication by using a load estimation algorithm \cite{HybridRAandDataM2M}. The details of the radio resource manager is out of the scope of this paper. In fact, our aim in this paper is to maximize the number of M2M devices which can be supported by a given number of radio resources and the proposed scheme can be combined with any dynamic resource management scheme to optimize the system-wide performance. Moreover, H2H users have high priority to obtain a connection to transmit their data to the BS. Most research to date has considered contention-free random access for H2H users, that is the BS assigns a preamble and a data channel to the H2H user in a timely manner \cite{PerfModelDelay}. In this work, we only focus on M2M devices, which opportunistically contend for data channels through a contention-based random access, and H2H users are assumed to have access to the BS through the allocated data channels. 

It is assumed that the channel between each M2M device and the BS is a slow time-varying block fading channel, for which the channel remains constant within one transmission block but varies slowly from one block to the other. We consider a time division duplex (TDD)-based wireless access system, where the channel gain of the uplink is assumed to be the same as that of the downlink \cite{OpPowerAlloc}. With this assumption, each device can estimate the uplink channel gain from the pilot signal sent periodically over the downlink channel by the BS. The BS, however, does not have knowledge of any channel state information (CSI). This assumption is particularly relevant in M2M communications with fixed location devices where, due to a large number of devices, it would be impractical for the BS to obtain CSI to every MTC device \cite{RayCSI}. Moreover, we assume that the devices perform power control in such a way that the received power from all the devices at the BS is the same.
\section{The Proposed Random Access Strategy for M2M communications}
In this section, a novel massive access strategy for M2M communication is proposed. Unlike the conventional RA strategy in LTE where only the existence of preambles are detected by the BS, in the proposed scheme the BS can effectively detect the preambles and estimate the number of devices which have selected each preamble using conventional Zadoff-Chu sequences \cite{SpatialGroupRAM2M} as RA preambles.
\begin{table}[t]
\caption{Notation Summary}
\label{NotSum}
\centering
\scriptsize
\begin{tabular}{|p{1cm}|p{6cm}|}
\hline
\textbf{Notation}&\textbf{Description}\\
\hline
$W_\mathrm{s}$& Total bandwidth of an RB in Hz\\
\hline
$\tau_\mathrm{s}$&The duration of an RB\\
\hline
$T_\mathrm{f}$& Time frame duration\\
\hline
$\gamma$& The total received SNR at the BS\\
\hline
$\gamma_0$& Received SNR at the BS after power control per MTC device\\
\hline
$\gamma_{\max}$& Maximum total received power at the BS\\
\hline
$\gamma_{0,\max}$& Maximum received SNR at the BS from each MTC device\\
\hline
$\lambda$& Average number of active devices in an RB\\
\hline
$N$& Total number of devices\\
\hline
$N_\mathrm{s}$& Total Number of RA preambles\\
\hline
$N_\mathrm{t}$& Total number of timing groups\\
\hline
$N_\mathrm{ZC}$& Length of the ZC sequence\\
\hline
$\gamma_\mathrm{th}$& Threshold SNR in the load estimation algortihm\\
\hline
$k$& Payload size of each MTC device\\
\hline
$T_\mathrm{s}$& Basic time unit which is equal to 32.552 ns\\
\hline
$\tau$& minimum time difference between two timing groups\\
\hline
\end{tabular}
\end{table}
\normalsize
\subsection{The Contention-Based RA Phase}
We assume that MTC devices perform power control such that the signal transmitted by each MTC device is received at the BS with the same power $P_0$. The steps of the proposed RA strategy are as follows:
\begin{enumerate}
 \item \textbf{PRACH scheduling}: Before each time frame begins, the BS decides the number of RBs for a PRACH and broadcast the configuration of RBs for a PRACH via a downlink control channel. In this paper, we assume that the number of RBs for the PRACH of M2M and their configuration is fixed. More specifically, we assume $N_\mathrm{s}$ preambles are allocated for RA of MTC devices.
  \item \textbf{Preamble transmission}: Each MTC device which has data to transmit, randomly chooses a preamble out of $N_\mathrm{s}$ available preambles with equal probability. The chosen preamble is then sent to the BS via the PRACH. 
  \item \textbf{Preamble detection and data channel scheduling}: The BS detects all the preambles transmitted on the PRACH by the MTC devices and determines the total number of active MTC devices. Then the BS broadcasts the scheduling information along with the information about the weight coefficients to all devices via a downlink control channel in the form of a random access response (RAR). The details of this step will be discussed in the next subsection. 
 \end{enumerate}
 It is important to note that in the third step of the proposed RA strategy, the BS broadcasts the information regarding the weight coefficients to all the devices. The weight coefficients are used in the data transmission phase and are designed such that the BS can accommodate all the detected devices within the available resource channels. We will discuss different weight designs and their respective rate performances in Section V.

\subsection{Load Estimation Algorithm}
The load estimation algorithm runs on Step 3 of the RA procedure and aims to determine the number of devices which have selected each preamble and been received with the same delay at the BS.

\begin{figure}[!t]
\centering
\includegraphics[scale=0.42]{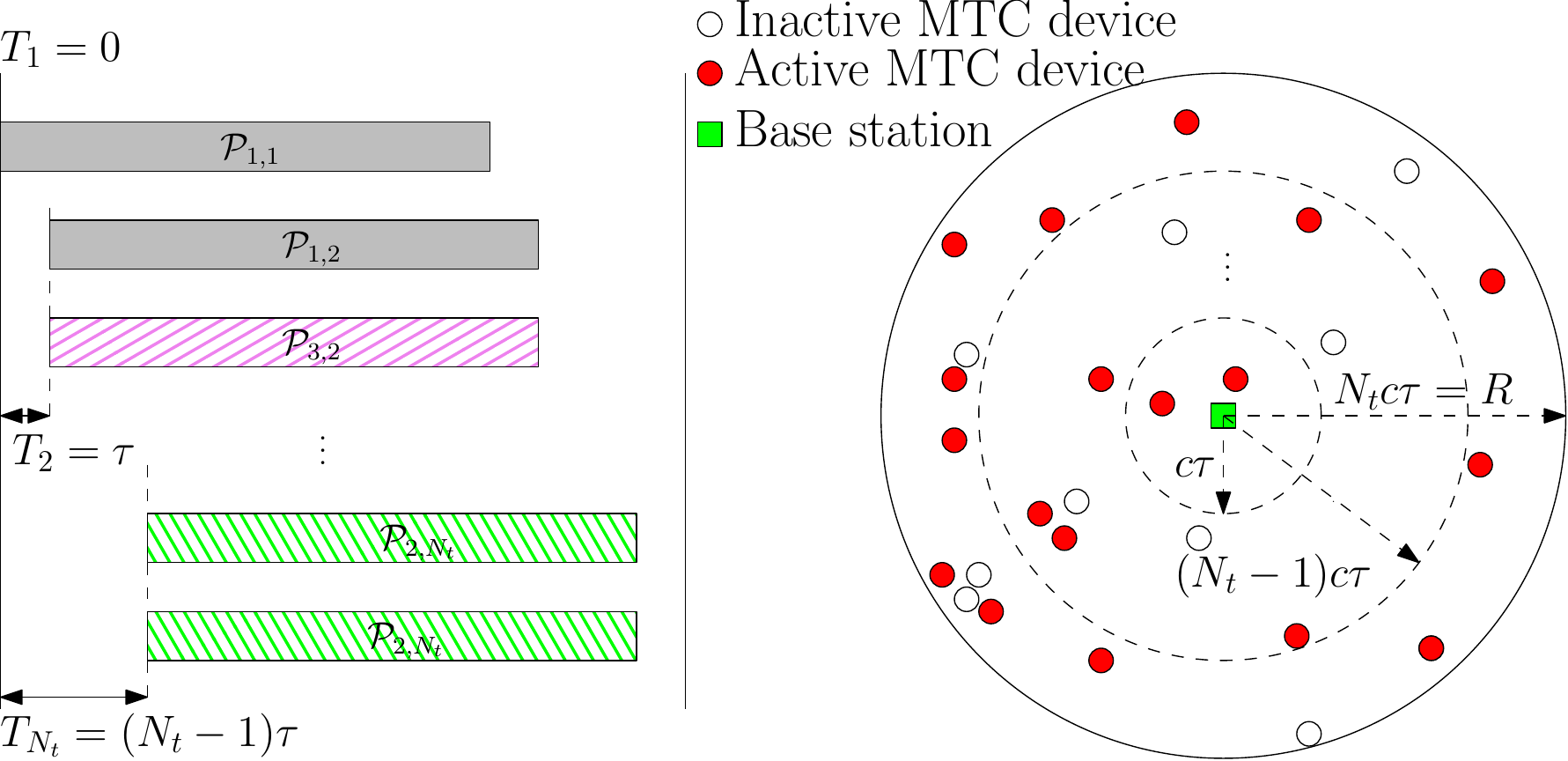}
\caption{Virtual cell partitioning based on timing advance information of MTC devices and the respective random access process in the RA phase.}
\label{timeadvance}
\end{figure}

In the conventional RA procedure in LTE, a set of information is sent as a RAR message in the third step of the RA phase. More specifically, the RAR message in LTE contains, 1) a number to identify the RA slot, 2) the index of the received preamble, 3) the timing advance command, and 4) the resource allocation information \cite{ACB_Time}. The timing advance command is used to adjust the uplink transmission time in such a way that the data is received at the BS at the anticipated time. This command takes an index value by a multiple of 16 $T_\mathrm{s}$, where $T_\mathrm{s}$ denotes the basic time unit and is equal to 32.552 ns \cite{ACB_Time}. Similar to \cite{ACB_Time}, we assume that two propagation delays are quantized to the same index when their difference is less than or equal to $\tau=8 T_\mathrm{s}$. Furthermore, the propagation delays of MTC devices to the BS are quantized and take values of multiple of $\tau$. Therefore, the propagation delay is modelled by an index taking values from $0$ to $N_\mathrm{t}$, where $N_\mathrm{t}$ is the maximum timing index which is determined by the cell coverage radius, $R$. More specifically, $N_\mathrm{t}=\lceil R/(c\tau)\rceil$, where $c=3\times 10^8$ m/s is the light speed and $\lceil.\rceil$ is the ceil operator. The cell coverage area can then be virtually partitioned into multiple regions, where the devices in each region have the same timing index. That is an MTC device at distance $r$ from the BS belongs to the $\ell^{th}$ timing group if $(\ell-1)\tau<r/c<\ell\tau$. Each MTC device determines its own group based on its distance information between the BS and itself, which can be obtained by several distance estimation algorithms \cite{SpatialGroupRAM2M}. Thus, we assume that each MTC device knows its own timing index. Fig. \ref{timeadvance} shows this cell partitioning based on timing indexes.

As shown in Fig. \ref{timeadvance}, the transmitted preambles by MTC devices are received with different delays (i.e., timing indexes) due to different propagation distances from MTC devices to the BS. In LTE, the BS determines the presence of any preamble by calculating the discrete cross correlation of the received signal with each of the $N_\mathrm{s}$ preambles.

The Zadoff-Chu (ZC) sequences are used to generate RA preambles which are defined as $z_r[n]=\exp[-j\pi r n(n+1)/N_{\mathrm{ZC}}]$ for $n=0, \cdots, N_{\mathrm{ZC}}-1$, where $N_{\mathrm{ZC}}$ is the sequence length and $r\in\{1, \cdots, N_{\mathrm{ZC}}-1\}$ is the root index \cite{SpatialGroupRAM2M}. The magnitude of the cyclic correlation of each ZC sequence with itself is a delta function, i.e., $c_{rr}[\sigma]=|\sum_{n=0}^{N_{ZC}}z_r[n]z^*[n+\sigma]|=N_{\mathrm{ZC}}\delta[\sigma]$, where $(.)^*$ denotes the complex conjugate. Using this property, we can determine by how much the received sequence is shifted. Multiple RA preambles are then generated from the ZC sequence by cyclically shifting the sequence by a factor of $N_{\mathrm{SC}}$, which is determined by the system parameters. The $i^{th}$ preamble can be generated as $z_{r,i}[n]=z_r[\mod(n+iN_{\mathrm{CS}},N_{\mathrm{ZC}})]$. More details on the ZC sequences and their design parameters in LTE can be found in \cite{ZCort,ZCnon,E-UTRA_Stand}.

\begin{figure}[t]
\centering
\includegraphics[scale=0.33]{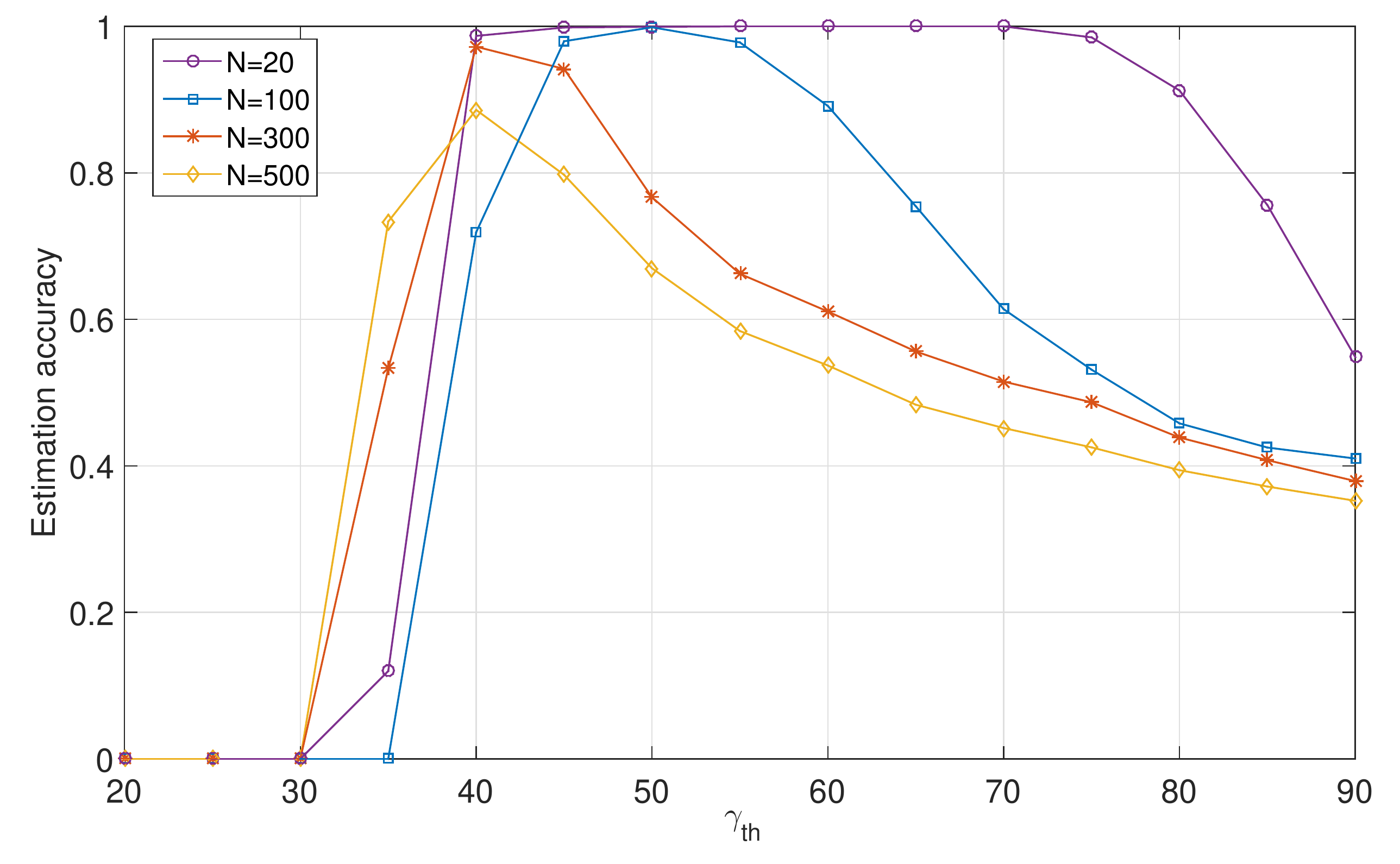}
\vskip-2ex
\caption{Estimation accuracy for different number of devices when $N_\mathrm{s}=N_\mathrm{t}=20$, $N_{\mathrm{ZC}}=100$, and $\mathrm{SNR} = 0$ dB.}
\label{estaccfig}
\end{figure}

Now, we explain the proposed load estimation algorithm based on timing advance and different power levels of received preambles. Let $n(i,j)$ denote the number of devices belong to the $j^{th}$ timing group that have selected the $i^{th}$ preamble. We denote by $\mathcal{P}_{i,j}$ the $i^{th}$ received preamble at the BS which is sent by the device in the $j^{th}$ timing group. This preamble is shifted by $(j-1)\tau$ due to the propagation delay of the $j^{th}$ group. The received signal at the BS at the end of the first step of the RA phase can be written as follows:
\begin{align}
Y=\sum_{i=1}^{N_\mathrm{s}}\sum_{j=1}^{N_\mathrm{t}} n(i,j)\mathcal{P}_{i,j} +Z,
\end{align}
where $Z$ is additive white Gaussian noise (AWGN) with zero mean and variance $\sigma^2_z$. The BS then calculates the cross-correlation between the received signal $Y$ and each preamble with different timing indexes. Unlike the preamble detection strategy in the LTE standard that only one timing advance is detected for multiple copies of each preamble received by the BS, the BS in the proposed scheme can detect all the timing indexes of all the copies of each preamble, thanks to the same received power from all the devices at the BS. Algorithm \ref{alg1} shows the steps of the proposed iterative load estimation strategy at the BS.
\begin{algorithm}
\caption{Load Estimation Algorithm}
\begin{algorithmic}[1]
\State {\textbf{Initialize} $\hat{\textbf{n}}=\textbf{0}$}
\While{$||Y||_2^2>\gamma_{\mathrm{th}}$}
\For{$i\in\{1,\cdots,N_\mathrm{t}\}$}
\For{$j\in\{1,\cdots,N_\mathrm{s}\}$}
\If{$|\sum_{n=0}^{N_{\mathrm{ZC}}} Y[n]\mathcal{P}_{j,i}[n]|>\gamma_{\mathrm{th}}$}
\State {$\hat{n}(i,j)++$,}
\State{$Y=Y-\mathcal{P}_{j,i}$,}
\EndIf
\EndFor
\EndFor
\EndWhile
\end{algorithmic}
\label{alg1}
\end{algorithm}
\vskip-3ex
It is important to note that $\gamma_{\mathrm{th}}$ in Algorithm 1 can be changed and optimized for different loads. Fig. \ref{estaccfig} shows the estimation accuracy of the proposed scheme under different loads. Here, the estimation accuracy is defined as the difference between the estimated and actual number of devices divided by the actual number of devices. As can be seen in this figure the proposed approach can accurately estimate the total number of devices from the preambles sent over the PRACH. The design and adjusting the parameters of the proposed algorithm can be done in such a way to maximize the accuracy of the estimation. This is however beyond the scope of the paper. In the rest of the paper, we assume that the BS always detects the preambles and accurately determines the number of devices which have selected each preamble and have the same timing index.

It worth noting that a similar algorithm is used in the current LTE standard to only detect each preamble. As several devices might have chosen each preamble the discrete cross correlation of the received sequence and a cyclically shifted preamble contains several impulses in the time domain. These impulses determine the propagation delay of the devices that has selected the respective preamble. However, the BS only consider the impulse with the shortest delay/highest amplitude, and broadcast the respective timing advance in the RAR message. This is inefficient as the BS only schedules a data channel for the device with the shortest propagation delay when multiple devices have selected the same preamble. Moreover, if more than one devices have the same timing advance and have selected the same preamble, their scheduled messages will collide and the respective data channel will be unused. 

\section{The Proposed Data Transmission Phase for M2M communications}
\subsection{An Overview on Raptor Codes in the Low SNR Regime}
A Raptor code \cite{Raptor} is a simple concatenation of a high-rate low density parity check (LDPC) code and a Luby transform (LT) code \cite{Luby}. A $k$-bit information sequence is first encoded by using a high-rate LDPC code to generate $k'$ LDPC coded symbols, also referred to as input symbols. Using an LT code, a potentially limitless number of coded (output) symbols can then be generated. The encoding process of LT codes contains two important steps. First, an integer $d$, called degree, is obtained from a predefined probability distribution function, called a degree distribution. Second, $d$ distinct input symbols are uniformly selected at random and then XORed to generate one output symbol. The encoding process will be terminated when the sender receives an acknowledgement from the destination or a pre-determined number of coded symbols are sent.

Let $\Omega_d$ denote the probability that the degree is $d$. Then, the degree distribution function can be represented in a polynomial form as follows:
\begin{align}
\Omega(x)=\sum_{d=1}^{D}\Omega_dx^d,
\label{Eq1}
\end{align}
where $D$ is the maximum code degree. 

A sum product algorithm (SPA) is usually used for the decoding of Raptor codes, where log-likelihood ratios (LLRs) are passed as messages along edges from variable to check nodes and vice versa in an iterative manner. More details of this decoder can be found in \cite{RaptorBSC}. The design of Raptor codes over AWGN channels in the low signal to noise ratio (SNR) regime has been studied in \cite{LowSNRRaotor_Mahyar}, where an exact expression for the degree distribution polynomial in the low SNR regime was found. More specifically, the asymptotic degree distribution polynomial in the low SNR regime when the maximum code degree goes to infinity is given by \cite{LowSNRRaotor_Mahyar}:
\begin{align}
\label{AsymDeg}
\Omega^{(\infty)}(x)=\frac{1}{4\ln(2)}\int_{0}^{x}\varphi^{-1}(t)dt,~~x\in[0,1],
\end{align}
where $\varphi^{-1}(x)$ is the inverse of $\varphi(x)$, which is defined as follows for $x>0$:
\begin{align}
\varphi(x)=\frac{1}{\sqrt{4\pi x}}\int_{-\infty}^{\infty}\tanh\left(\frac{u}{2}\right)\text{e}^{-\frac{(u-x)^2}{4x}}du.
\label{phifunc}
\end{align}
A set of practical degree distributions with limited maximum degree was also designed in \cite{LowSNRRaotor_Mahyar}, which have shown excellent rate performance in very low SNRs (below -10 dB). More specifically, a rate efficiency of 0.95 was achieved for a Raptor code with maximum degree 300 in the whole SNR range below -10 dB. Here, the rate efficiency is defined as the ratio of the achievable rate and the channel capacity. This is an interesting property for Raptor codes, where a single degree distribution can be used for all SNRs below -10 dB to achieve a near capacity performance over AWGN channels.
\subsection{Encoding at MTC devices}
In the data transmission phase, each MTC device appends its unique ID to its message and then encodes it by using a Raptor code. Each device uses its preamble index and timing index as the seed for its random generator in the Raptor encoder, thus the BS and the device can build the same generator matrices for their Raptor codes. The same degree distribution is used for all the devices. This will significantly reduce the system design complexity as the devices do not need to change their code structure every time according to the system load. Moreover, as the MTC devices are assumed to have small packets to transmit, the effective code rate of each device can be very small. This allows to use the degree distribution optimized for Raptor codes in the low SNR regime \cite{LowSNRRaotor_Mahyar}, for all the devices. All the active devices which have successfully received the RAR message in the RA phase are transmitting at the same data channel, which consists of multiple RBs determined by the BS in the RA phase.
\begin{figure}[t]
\centering
\includegraphics[scale=0.46]{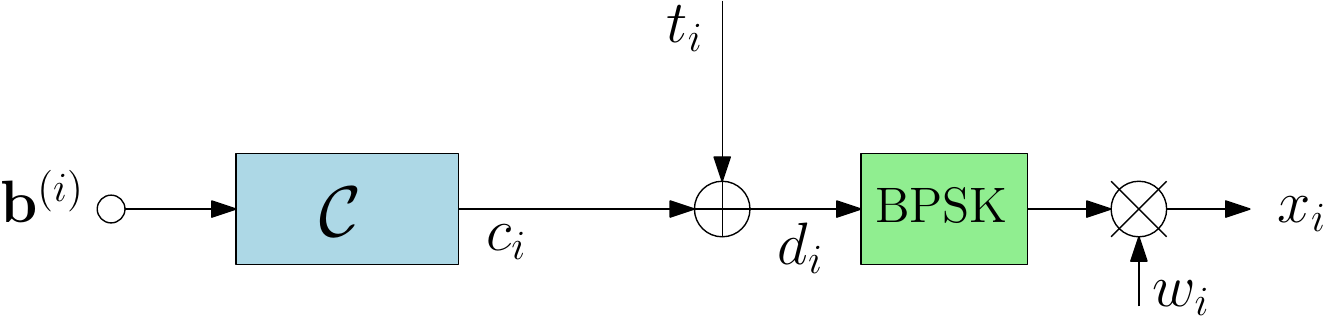}
\caption{Encoder structure at the $i^{th}$ MTC device with Raptor component code $\mathcal{C}$. $t_i$ is an i.i.d. random binary sequence and $w_i$ is the weight coefficient selected by the $i^{th}$ device.}
\label{EncDev}
\end{figure}

Fig. \ref{EncDev} shows the encoder structure at each MTC device. As can be seen in this figure, each encoded symbol $c_i$ is XORed with a binary random symbol $t_i$. We refer to this binary random source as the channel adaptor as it forces the symmetry condition for each equivalent binary input AWGN channel of each device. The resultant symbol, $d_i$, is then BPSK modulated and multiplied by the selected weight and transmitted over the scheduled data channel. The seed for the random generators of the channel adaptors at the devices are shared with the BS in the RA phase through the RAR message or at the beginning of the data transmission phase. This will be discussed in Section VI. The details of the weight coefficient design will also be discussed in the next section.
\begin{figure*}[t]
\centering
\includegraphics[scale=0.41]{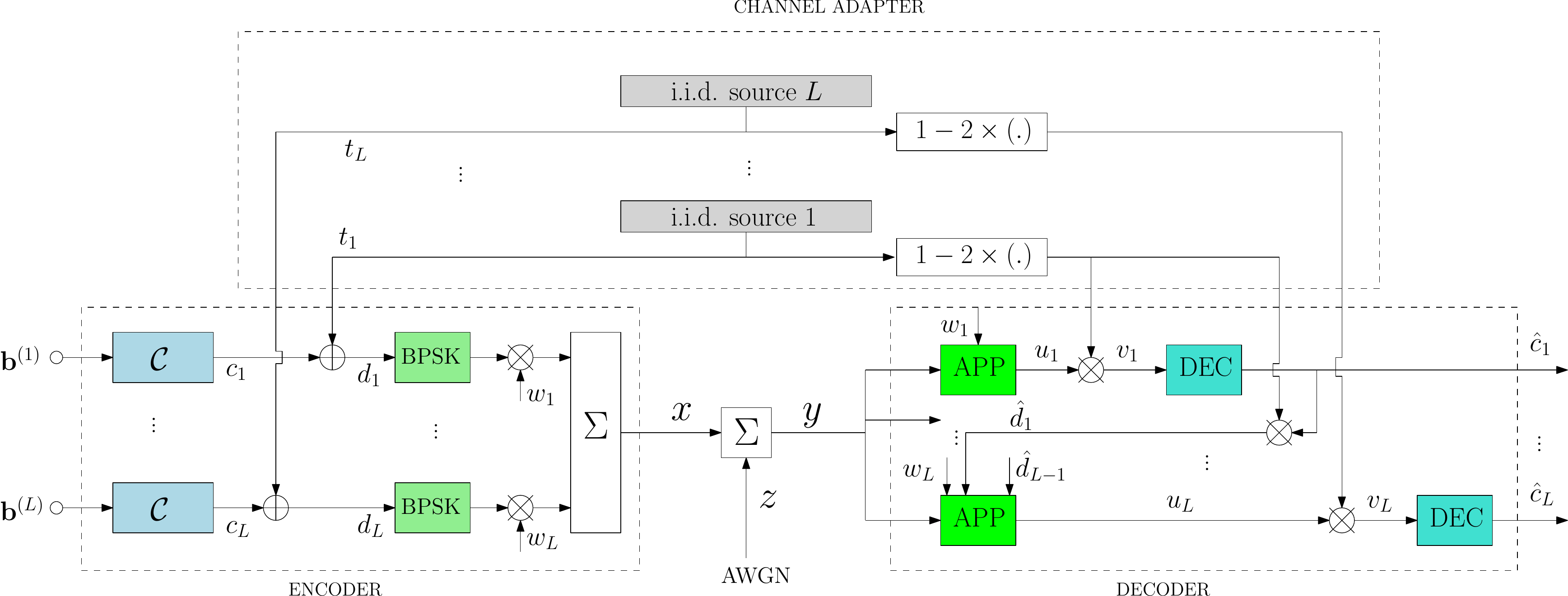}
\caption{Encoder and decoder structure of the proposed code with Raptor component codes.}
\label{EncUniv}
\end{figure*}
\subsection{Decoding at the BS}
Let $x_i$ denote the output of the BPSK modulator of the $i^{th}$ device, where we ignore the time index for the simplicity of representation. Then, the received signal at the BS, denoted by $y$, is shown as follows:
\begin{align}
y=\sum_{i=1}^{L}w_ix_i+z,
\label{origchan}
\end{align}
where $L$ is the number of active MTC devices, $z$ is  zero mean additive white Gaussian noise with variance $\sigma^2_z$. Fig. \ref{EncUniv} shows the encoder structure of the multi-layer realization of the proposed scheme, where each layer corresponds to one MTC device transmitting at the same data channel. For the decoding of the devices' messages, we use the well-known multistage decoding (MSD). More specifically, the decoder for the $i^{th}$ stage, i.e., a decoder for the $i^{th}$ MTC device at the BS, removes all coded symbols from all previous stages, and treats the coded symbols of stages $i+1$ to $L$ as additive noise. Let $y_i$ denote the effective input of the $i^{th}$ stage of the decoder, it can be shown as follows:
\begin{align}
y_i=w_ix_i+z_i,
\label{equichani}
\end{align}
where $z_i=\sum_{j>i}w_jx_j+z$ is the effective noise of the $i^{th}$ decoder stage. It can be easily verified that $z_i$ has zero mean and its variance, denoted by $\sigma_i^2$ can be calculated as follows:
\begin{align}
\sigma^2_i=\sum_{j=i+1}^{L}w_j^2+\sigma^2_z.
\end{align}
The input of the $i^{th}$ stage of the decoder can then be realized as the output of an equivalent BI-AWGN channel, where its effective SNR is given by:
\begin{align}
\gamma_i=\frac{w_i^2}{\sum_{j=i+1}^L w_j^2+\sigma^2_z}.
\label{equivsnr}
\end{align}

It is important to note that by using i.i.d. channel adapters, we can force the symmetry of the equivalent binary-input component channels \cite{CM_LDPC_IT}. Generally, a binary input channel is symmetric if \cite{CM_LDPC_IT}
\begin{align}
\nonumber p(Y=y|C=0)=p(Y=-y|C=1)
\end{align}
where $c$ and $y$ are input and output of the binary-input channel, respectively.  As shown in Fig. \ref{EncUniv}, the sign of the channel output will be adjusted by $v_i=u_i(1-2t_i)$, where $u_i$ is the LLR of the APP module output, and $v_i$ is the input of the $i^{th}$ stage decoder. More specifically, the APP module at stage $i$ will subtract the coded symbols of the previous stages from the channel output $y$, and calculate the output LLR, $u_i$, of the equivalent BI-AWGN channel (\ref{equichani}) as follows:
\begin{align}
\mathrm{LLR}_i=\frac{2w^2_i y_i}{\sigma^2_i}.
\end{align}
As shown in \cite{CM_LDPC_IT}, for any new binary-input output symmetric component channel, if we use a channel code $\mathcal{C}$ through this channel, the decoding error probability is independent of the codeword. It was also shown that the capacity of the binary input channel with i.i.d. equiprobable input distribution is equal to the capacity of the equivalent binary-input output symmetric channel with the i.i.d. channel adapter. This means that no rate loss is introduced in the system due to use of the channel adapter.
\section{Analysis of of the Proposed Scheme and Weight Coefficient Design}
\subsection{Achievable Rate}
With the well-known multi-stage decoding and the mutual information chain rule, we have
\begin{align}
\nonumber I(C_1,\cdots,C_L;Y)=&~I(C_1;Y)+I(C_2;Y|C_1)+\cdots\\
&+I(C_L;Y|C1,\cdots,C_{L-1}),
\label{chainrule}
\end{align}
which shows that the transmission of vector $(c_1,\cdots,c_L)$ can be separated into the parallel transmission of $c_i$ over $L$ equivalent binary input channels \cite{CM_LDPC_IT}. More specifically, the mutual information for the $i^{th}$ equivalent binary input channel can be shown as follows:
\begin{align}
I(C_i;Y|C1,\cdots,C_{i-1})\approx\log_2(1+\gamma_i),
\label{appmuti}
\end{align}
which is valid since we assume that the equivalent SNR for all layers is very small. By using (\ref{equivsnr}), (\ref{chainrule}) can be rewritten as follows:
\begin{align}
\nonumber I(C_1,\cdots,C_L;Y)&=\sum_{i=1}^{L}\log_2\left(1+\frac{w_i^2}{\sum_{j=i+1}^L w_j^2+\sigma^2_\mathrm{z}}\right)\\
\nonumber &=\log_2\left(\prod_{i=1}^{L}\frac{\sum_{j=i}^L w_j^2+\sigma^2_\mathrm{z}}{\sum_{j=i+1}^L w_j^2+\sigma^2_\mathrm{z}}\right)\\
&=\log_2\left(\frac{\sum_{j=1}^L w_j^2+\sigma^2_\mathrm{z}}{\sigma^2_z}\right)=\log_2\left(1+\gamma\right),
\label{chainrule2}
\end{align}
which can be further simplified according to (\ref{appmuti}):
\begin{align}
\log_2(1+\gamma)=\sum_{i=1}^L \log_2(1+\gamma_i).
\label{equivrate}
\end{align}
As we assume that the devices continue their transmission over all the allocated RBs, i.e., the BS does not send acknowledgement to each individual device upon successful decoding of each stage, the achievable rate for each device is upper bounded by the rate of the device with the minimum effective SNR. Let $R_i$ denote the effective rate of each device, then we have:
\begin{align}
R_i\le \log_2(1+\gamma_{\min}),
\end{align}
where $\gamma_{\min}=\min_i \{\gamma_i\}$ and $R_{\min}\triangleq \log_2(1+\gamma_{\min})$.

Let $k$ denote the message length of each MTC device, including the device ID. The number of RBs required for the successful transmission of the message can be calculated as follows:
\begin{align}
V\ge\left\lceil \frac{k}{\tau_sW_sR_{\min}}\right\rceil.
\end{align}
The RB load, defined as the average number of devices per RB, can be easily obtained as $V/N$.
\subsection{Designing the Weight Coefficients}
As can be seen in (\ref{equivrate}), the effective rate for each device is characterized by the weight coefficients. In this section, we propose three weight coefficient designs and discuss their respective rate performances.
\subsubsection{Equal Weight Selection}
In the equal weight (EqW) selection strategy, all the devices select the same weight coefficient. Let $w$ denote the weight coefficient selected by all devices, then, it can be calculated as follows:
\begin{align}
w=\frac{1}{\sqrt{L}},
\end{align}
which is due to the fact that the overall received signal power at the BS is assumed to be 1. The effective SNR of the $i^{th}$ device can then be calculated as follows:
\begin{align}
\gamma_i=\frac{w^2}{(L-i)w^2+\sigma^2_\mathrm{z}}=\frac{1}{L-i+L\sigma^2_\mathrm{z}},
\end{align}
and the minimum SNR for MTC devices is given by
\begin{align}
\gamma_{\min}=\frac{1}{L-1+L\sigma^2_\mathrm{z}}.
\end{align}
The maximum achievable rate for MTC devices is then upper bounded by the rate of the device with the minimum SNR. It can be characterized as follows:
\begin{align}
R^{(\mathrm{EqW})}_{\min}=\log_2\left(1+\frac{1}{L-1+L\sigma^2_\mathrm{z}}\right).
\end{align}

\subsubsection{Exponential Weight Selection}
In the exponential weight (ExW) selection strategy, the weight coefficients are designed in such a way that the effective SNR at the BS for each device is the same.
\begin{lemma}
\label{OptWeiLemma}
Let $L$ denote the number of active MTC devices and $\gamma_0$ denote the target effective SNR at the BS for every  MTC device. Then, the overall received SNR at the BS is given by:
\begin{align}
\gamma=\left(1+\gamma_0\right)^L-1,
\label{numberlayers}
\end{align}
and the optimal weight coefficients can be calculated as follows, for $i=0, \cdots, L-1$:
\begin{align}
w_{L-i}^*=\sqrt{(1+\gamma_0)^i\frac{\gamma_0}{\gamma}}.
\label{weightfor}
\end{align}
\end{lemma}
\begin{IEEEproof}
Since the weight coefficients are designed such that the effective SNR of each equivalent BI-AWGN channel is $\gamma_0$, according to (\ref{equivrate}) we have:
\begin{align}
L \log_2(1+\gamma_0)=\log_2(1+\gamma),
\end{align}
which directly results in (\ref{numberlayers}). Also from (\ref{equivsnr}), it is clear that for the $L^{th}$ layer, we have $w_L^2=\gamma_0/\gamma$, which proves (\ref{weightfor}) for $i=0$. We assume that (\ref{weightfor}) holds for $i=0,\cdots, j$ for $j\ge 0$, we then show that it also holds for $j+1$. As we assume that the effective SNR for all layers is $\gamma_0$, then for layer $L-j-1$ we have:
\begin{align}
\nonumber\gamma_0&=\frac{w_{L-j-1}^2}{\sum_{i=L-j}^{L}w_i^2+\sigma^2_\mathrm{z}}=\frac{w_{L-j-1}^2}{\sum_{i=0}^{j}w_{L-i}^2+\sigma^2_\mathrm{z}}\\
\nonumber &=\frac{w_{L-j-1}^2}{\sum_{i=0}^{j}(1+\gamma_0)^i\frac{\gamma_0}{\gamma}+\sigma^2_\mathrm{z}}\\
\nonumber &=\frac{\gamma w_{L-j-1}^2}{(1+\gamma_0)^{j+1}},
\end{align}
which is equivalent to (\ref{weightfor}) for $j+1$. This completes the proof. 
\end{IEEEproof}

In this scheme, the devices randomly select the weight coefficients from the set of optimal weight coefficients. Therefore, it is highly probable that more than one device will select the same weight coefficient. Let $r_i$ denote the number of devices which have selected the weight coefficient $w^*_i$. It is clear that $\sum_{i=1}^L r_i=L$ and $w_i=w^*_j$ for $1+\sum_{\ell=1}^{j-1}n_{\ell}\le i\le\sum_{\ell=1}^{j}n_{\ell}$. According to (\ref{equivsnr}), it is easy to verify that $\gamma_i<\gamma_j$ for every $i<j$ where $1+\sum_{\ell=1}^{j-1}n_{\ell}\le i, j \le\sum_{\ell=1}^{j}n_{\ell}$. This is because
\begin{align}
\nonumber \gamma_i&=\frac{w_i^2}{\sum_{j=i+1}^L w_j^2+\sigma^2_\mathrm{z}}=\frac{w^{*2}_i}{\sum_{j=i+1}^L w_j^2+\sigma^2_\mathrm{z}}\\
&<\frac{w_j^2}{\sum_{\ell=j+1}^L w_{\ell}^2+\sigma^2_\mathrm{z}}=\gamma_j,
\label{uneq1}
\end{align}
which is valid when $i<j$. Accordingly, the minimum effective SNR for MTC devices can be calculated as follows:
\begin{align}
\gamma_{\mathrm{min}}= \min_{\{i: r_i\ne0\}} \frac{w_{i}^{*2}}{(r_i-1)w_{i}^{*2}+\sum_{j={i}+1}^L r_j w_{j}^{*2}+\sigma^2_\mathrm{z}}.
\end{align}

In the following, we find the probability distribution function (pdf) of $\gamma_{\min}$. For this aim, we define $\xi_i=\frac{w_{i}^{*2}}{(r_i-1)w_{i}^{*2}+\sum_{j={i}+1}^L r_j w_{j}^{*2}+\sigma^2_\mathrm{z}}$ for $i\in\{i:r_i\ne 0\}$; otherwise it is set to $\infty$. Then according to Lemma \ref{OptWeiLemma} and (\ref{weightfor}), we have:
\begin{align}
\nonumber \frac{1}{\xi_i}=(r_i-1)+\sum_{j=i+1}^L r_j(1+\gamma_0)^{i-j}+\frac{1}{\gamma_0}(1+\gamma_0)^{i-L}.
\end{align}
Since the devices randomly choose from the $L$ available weight coefficients, for a sufficiently large $L$, $r_i$ can be modelled by a binomial distribution with a success probability of $1/L$, i.e.,
\begin{align}
p(r_i)=\dbinom{L}{r_i}(1/L)^{r_i}(1-1/L)^{L-r_i},
\end{align}
and its mean and variance are $\mathbb{E}[r_i]=1$ and $\mathrm{var}[r_i]=(1-1/L)$, respectively. It is also easy to show that $p(r_i=0)=(1-1/N)^N$. When $L$ goes to infinity and $r_i\ne0$, $1/\xi_i$ can be modelled by a  Gaussian distribution according to the Central Limit Theorem, with mean $m_i$ and variance $S_i$, which are given by:
\begin{align}
\nonumber m_i=\mathbb{E}[1/\xi_i]&=\sum_{j=i+1}^L r_j(1+\gamma_0)^{i-j}+\frac{1}{\gamma_0}(1+\gamma_0)^{i-L}=\frac{1}{\gamma_0},\\
\nonumber S_i=\mathrm{var}[1/\xi_i]&=(1-\frac{1}{L})\sum_{j=i}^L (1+\gamma_0)^{2(i-j)}.
\end{align}
The complete pdf of $1/\xi_i$ can then be characterized as follows:
\[p\left(\frac{1}{\xi_i}=z\right)=\left\{\begin{array}{ll}
p_0;&z=0,\\
\frac{1-p_0}{\sqrt{2\pi S_i}}\exp\left(-\frac{(z-m_i)^2}{2S_i}\right);&z\ne0,\end{array}\right.\]
where $p_0=(1-1/N)^N$. The cumulative distribution function (cdf) of $\gamma_{\min}$, denoted by $F_{\gamma}(x|L)$, when the number of devices is $L$, can be found as follows:
\begin{align}
\nonumber F_{\gamma}(x|L)&=p(\gamma_{\min}<x)=p(\min_i{\xi_i}<x)=p(\max_i \frac{1}{\xi_i} >\frac{1}{x}) \\
\nonumber &= 1-p(\max_i \frac{1}{\xi_i} <\frac{1}{x})=1-p(\frac{1}{\xi_1} <\frac{1}{x},\cdots, \frac{1}{\xi_L} <\frac{1}{x})\\
\nonumber &=1-\prod_{i=1}^{L} p(\frac{1}{\xi_i} <\frac{1}{x})\\
&= 1-\prod_{i=1}^{L} \left( p_0+(1-p_0)\left(1-Q\left(\frac{\frac{1}{x}-m_i}{\sqrt{S_i}}\right)\right)\right),
\label{approx1}
\end{align}
where $Q(x)=\frac{1}{\sqrt{2\pi}}\int_{x}^{\infty}\exp\left(-u^2/2\right)du$. As $L$ follows a Poisson distribution with mean $\lambda$, the complete cdf of $\gamma_{\min}$ is calculated as follows:
\begin{align}
F_{\gamma}(x)=\sum_{j}e^{-\lambda}\frac{\lambda^j}{j!} F_{\gamma}(x|L).
\label{approx2}
\end{align}
The pdf of $\gamma_{\min}$ is simply the derivative of $F_{\gamma}(x)$.
\begin{figure}[t]
\centering
\includegraphics[scale=0.28]{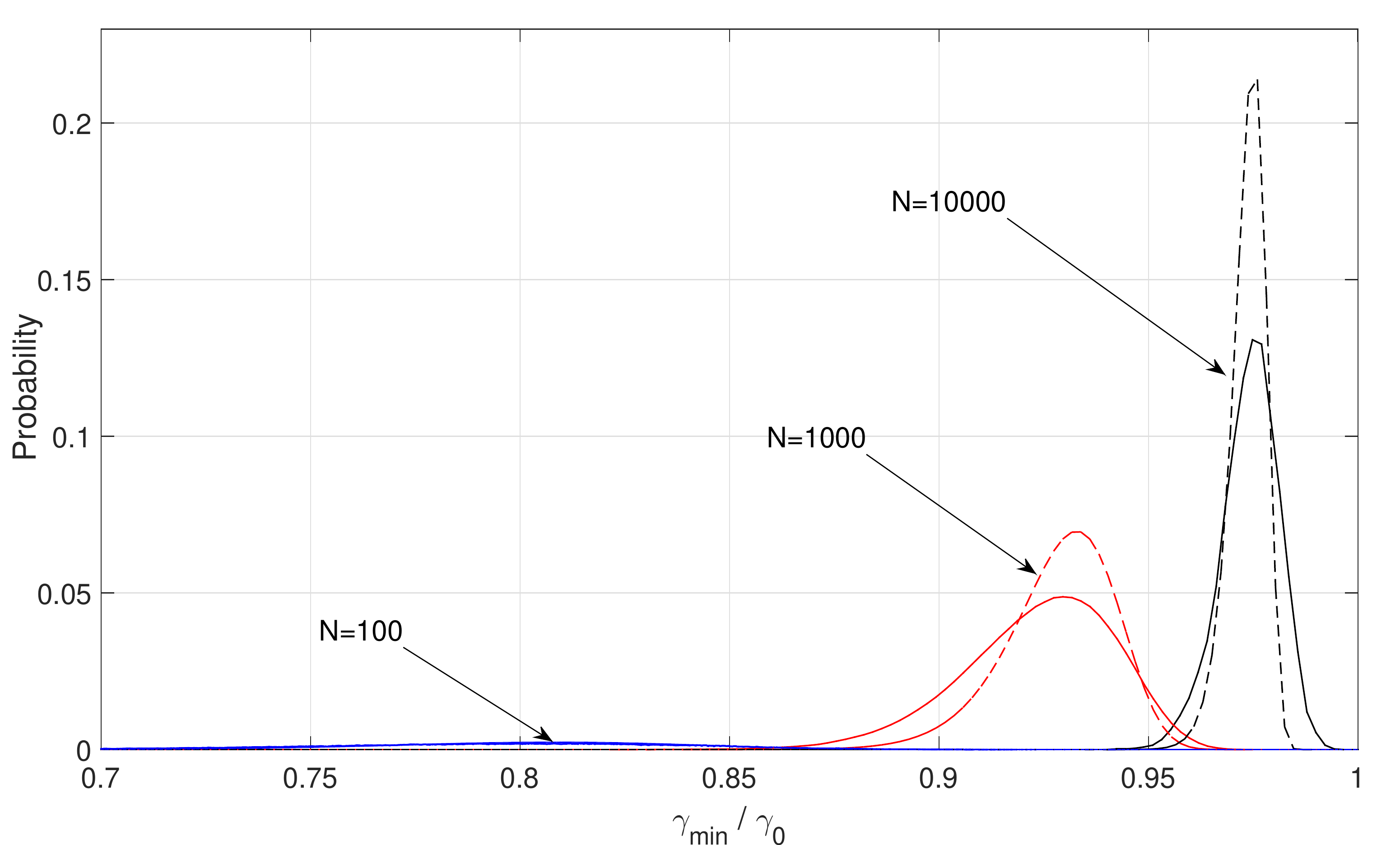}
\vskip-2ex
\caption{Histogram of the minimum device SNR for different number of devices, when the total SNR at the BS is $\gamma = 20$ dB. Solid lines shows the simulation results and dashed lines show the approximated results by using (\ref{approx1}).}
\label{HistMinSNRfig}
\end{figure}

Fig. \ref{HistMinSNRfig} shows the histogram of the minimum SNR divided by the target effective SNR, $\gamma_0$, when the total SNR at the BS is $\gamma = 20$ dB. As can be seen in this figure, with increasing the number of devices, and accordingly the number of available weight coefficients, the probability that the minimum SNR is closer to $\gamma_0$ is increased. This is because by increasing the number of devices, the probability that the devices choose separate weight coefficients is increased, which accordingly leads to a higher minimum achievable rate. It is important to note that the proposed approximation of the pdf shows around 1\% to 5\% mismatch in the mean value to the simulated pdf. This figure also shows the approximation of the pdf of $\gamma_{\min}$ by shifting the mean value by about 3\%, which is in a close agreement with the simulation results.

\begin{figure}[t]
\centering
\includegraphics[scale=0.28]{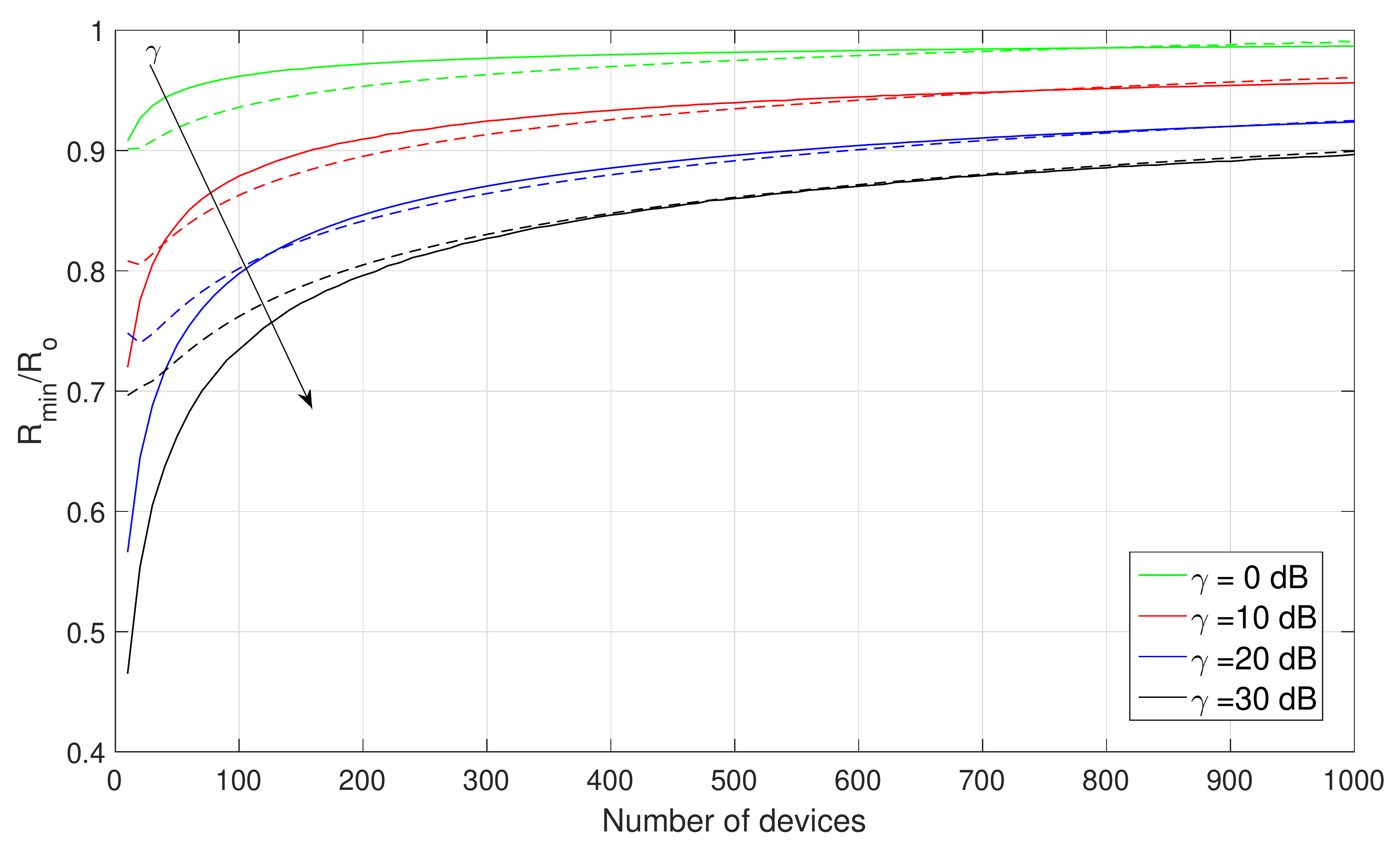}
\vskip-2ex
\caption{Average minimum achievable rate of MTC devices versus the number of devices for different target SNRs at the BS. Solid lines shows the simulation results and dashed lines show the approximated pdf using (\ref{approx1}).}
\label{RminNfig}
\end{figure}

Fig. \ref{RminNfig} shows the minimum achievable rate for MTC devices versus the number of devices. As can be seen in this figure, with increasing the number of devices, the minimum achievable rate gets closer to the designed rate. Moreover, with decreasing the target SNR at the BS, the minimum achievable rate gets closer to the desired rate. This is because, according to Lemma (\ref{OptWeiLemma}), with decreasing total SNR at the BS, the effective target SNR per device, $\gamma_0$, is decreased. More specifically, the ratio of weight coefficients for the $i^{th}$ and $j^{th}$ device, which is given by $\sqrt{(1+\gamma_0)^{i-j}}$, gets closer to 1. This leads to less rate loss as the weight coefficients are now close to each another. It is also important to note that the approximation of the average minimum achievable rate using (\ref{approx1}) shows excellent agreement with simulation results, especially when the number of devices is relatively large.

\subsubsection{Grouped Weight Selection}
Let us assume that the BS can determine the number of devices which have a particular timing advance and have selected a particular preamble. This way the devices can be partitioned into $N_{\mathrm{st}}=N_\mathrm{s}\times N_\mathrm{t}$ different groups, where the devices in each group have the same timing index and have selected the same preamble. The devices in each group are then assumed to have the same weight coefficient. We refer to this strategy as the group weight (GrW) selection strategy. The weight coefficients are then designed in such a way that the minimum effective SNR for the devices is maximized. Let $r_i$ denote the number of devices in the $i^{th}$ group, where $i=1,\cdots, N_{\mathrm{st}}$. We define the following optimization problem to maximize the minimum effective SNR for MTC devices:
\begin{align}
\nonumber \max_{\mathcal{W}} \min_{\{i:r_i\ne 0\}}&\frac{w_i^{2}}{(r_i-1)w_i^2+\sum_{j=i+1}^{N_{\mathrm{st}}}r_jw_j^2+\sigma^2_\mathrm{z}},\\
\nonumber \text{s.t.}~&(i)~\sum_{i=1}^{N_{\mathrm{st}}}r_iw_i^2=1,\\
\nonumber &(ii)~0\le w_i\le1, ~~~\mathrm{for}~ i=1,\cdots,N_{\mathrm{st}},
\label{optprob}
\end{align}
where we assume that the BS always starts the decoding from the group with the largest weight coefficient. To further simplify the design of the weight coefficients, we define a target minimum SNR $\gamma_0$, and determine the weight coefficients such that the minimum device SNR is at least $\gamma_0$. For this aim, we find the weight coefficients such that the effective SNR of the first device of each group is at least $\gamma_0$. This is because the first device of each group has the lowest effective SNR amongst other devices in the group, which can be easily proven by using the same strategy as in (\ref{uneq1}). Therefore, for the first device of the $i^{th}$ group, we have:
\begin{align}
\gamma_i=\frac{w_i^2}{(r_i-1)w_i^2+\sum_{j=i+1}^{N_{\mathrm{st}}}r_jw_j^2+\sigma^2_\mathrm{z}}=\gamma_0.
\end{align}
For $i=N_{\mathrm{st}}$, we have:
\begin{align}
\gamma_0=\frac{w_{N_{\mathrm{st}}}^2}{(r_{N_{\mathrm{st}}}-1)w_{N_{\mathrm{st}}}^2+\sigma^2_\mathrm{z}},
\end{align}
and the weight coefficient for the $N_{\mathrm{st}}^{th}$ group, i.e., $w_{N_{\mathrm{st}}}$, can be calculated as follows:
\begin{align}
w_{N_{\mathrm{st}}}^2=\frac{\gamma_0}{\gamma}\frac{1}{1-\gamma_0(r_{N_{\mathrm{st}}}-1)}.
\end{align}
The weight coefficient for the $i^{th}$ group can then be calculated as follows:
\begin{align}
w_i^2=\frac{\gamma_0(1+\gamma_0)^{{N_{\mathrm{st}}}-i}}{\gamma}\prod_{\ell=i}^{{N_{\mathrm{st}}}}\frac{1}{1-\gamma_0(r_{\ell}-1)}.
\label{groupweight}
\end{align}
As we assume that the total receive power at the BS is 1, then we have:
\begin{align}
\sum_{i=1}^{{N_{\mathrm{st}}}}r_iw_i^2=1,
\end{align}
and by substituting (\ref{groupweight}), we have:
\begin{align}
\gamma=\gamma_0\sum_{i=1}^{{N_{\mathrm{st}}}}\frac{r_i(1+\gamma_0)^{{N_{\mathrm{st}}}-i}}{\displaystyle\prod_{\ell=i}^{{N_{\mathrm{st}}}}{\left[1-\gamma_0(r_i-1)\right]}}.
\label{GrWfor}
\end{align}
It is clear from (\ref{GrWfor}) that for a given $\gamma_0$, the order of the number of devices in each group affects the total received SNR at the BS. The BS then performs an optimization to find the optimal order of the weight coefficients to maximize the average received SNR at the BS. For simplicity, we assume that $r_i\ge r_j$ for $i>j$. 
\subsection{Comparison Between Weight Selection Strategies}
 \begin{figure}[t]
\centering
\includegraphics[scale=0.28]{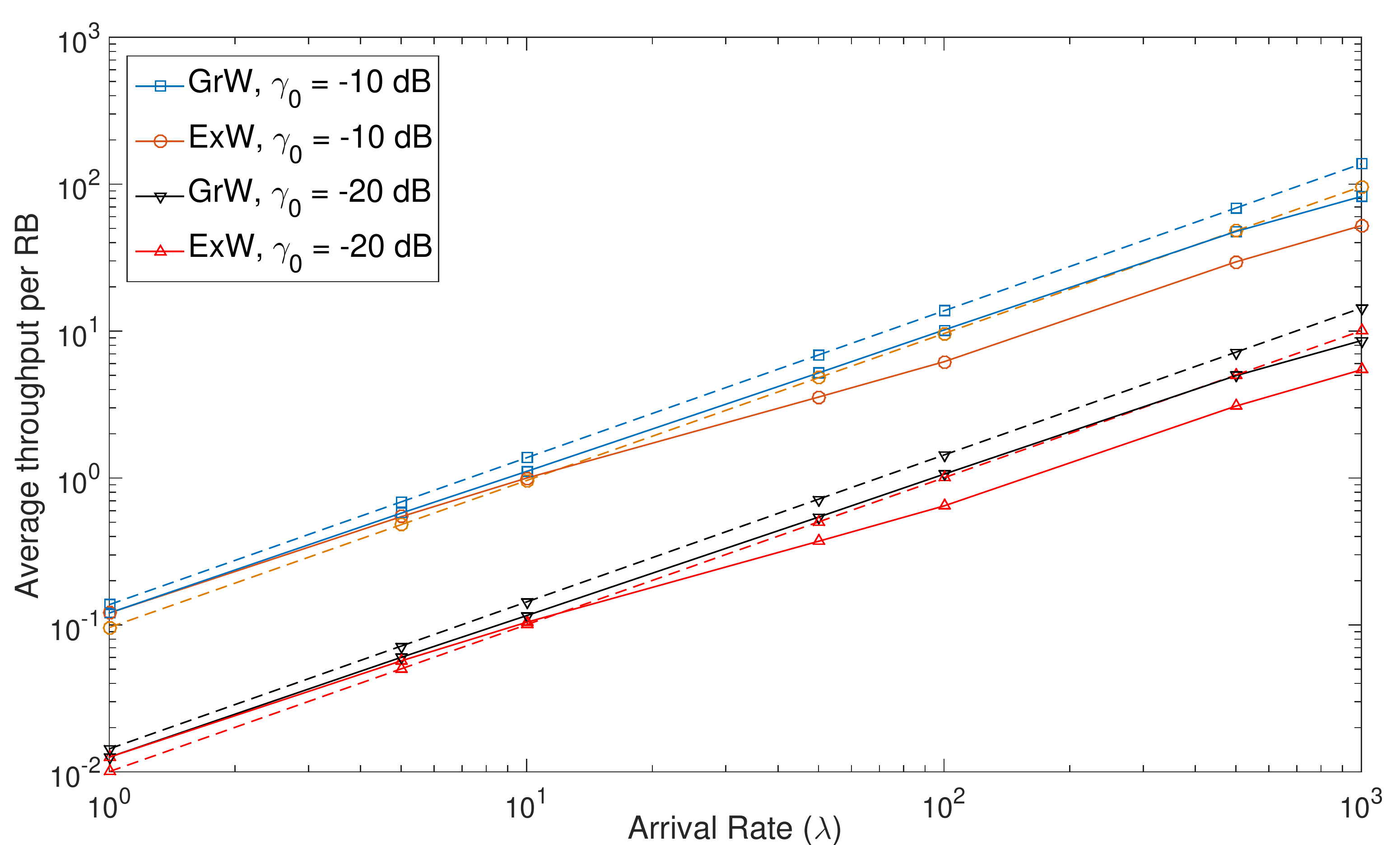}
\vskip-2ex
\caption{Average throughput per RBs for the proposed scheme with different weighting strategies without adaptive power management at the devices. The message length of MTC devices is $k=1024$, $N_\mathrm{s}=N_\mathrm{t}=20$, $W_\mathrm{s}=1$ MHz, and $\tau_{\mathrm{s}}=1$ ms. Solid and dashed lines show simulation and analytical results, respectively. }
\label{Sim1Fig}
\end{figure}
Fig. \ref{Sim1Fig} shows the RB load versus the number of devices for different target SNR values at the BS.  Fig. \ref{Sim1Fig} shows the average throughput per RB for the proposed scheme with different weighting strategies. The analytical results are also plotted in Fig. \ref{Sim1Fig} which show a good match with the simulation results. As can be seen in this figure, the grouped based weight selection can accommodate more devices per RB compared to the exponential weight selection strategy. This is because of non-ideal weight selection in the ExW scheme, which reduces  $\gamma_{\min}$ and accordingly the minimum achievable rate per MTC device decreases. However, in the GrW scheme, the weight coefficients are designed to maximize the lowest effective SNR, which can significantly increase the achievable rate per MTC device, especially in higher loads.

Moreover, as can be seen in Fig. \ref{Sim1Fig}, with increasing the arrival rate the number of MTC devices which can be supported by each RB is linearly increasing. This is due to the fact that the weight selection strategies are designed in such a way that the minimum received SNR per device remains constant at the BS. Thus, by increasing the number of devices the minimum achievable rate remains constant, therefore the BS can support a larger number of devices per RB.
\section{Practical Considerations}
\subsection{Overhead due to the Weight Selection in the Data Transmission Phase}
In the data transmission phase, the devices should select the weight coefficients and the BS must know which weights are selected and how many devices have selected the same weight. The devices then need to inform the BS of their selected weight coefficients. For this aim, we assume that devices randomly select a weight coefficient among $N$ coefficients and send the index of the weight to the BS. This can be done by sending a binary sequence of an appropriate length to the BS. The selected sequences should be orthogonal, so the BS can detect them. The information about these orthogonal sequences can be sent via the RAR message. Then a similar strategy as Algorithm 1 can be performed to detect the sequences and estimate the number of devices which have selected the same sequence. More specifically, we assume that the length of the weight sequences is $\delta N$, where $\delta\ge1$. $\delta$ is determined such that the probability that the BS cannot detect any sequence is minimized. The design of binary orthogonal sequences for synchronous and asynchronous CDMA systems has been widely studied and we do not discuss them here as it is out of scope of this paper. Interested readers are referred to \cite{ZCort} and the references therein for further details.

The devices perform timing adjustment and power control to transmit the selected sequences to the BS, thus the BS performs Algorithm 1 with a slight modification that the sequences are received without delay. This however increases the amount of overhead for the proposed scheme, as extra RBs have to be allocated for transmitting orthogonal weight sequences. The number of RBs required to successfully transmit orthogonal weight sequences in the data  transmission phase can be characterized as follows:
\begin{align}
N_{\mathrm{RB}}=\left\lceil\frac{\delta N}{R_\mathrm{w}W_\mathrm{s}\tau_\mathrm{s}}\right\rceil,
\end{align}
where $R_\mathrm{w}=\log_2(1+\gamma_\mathrm{w})$ and we assume that the devices perform power control and the received SNR for each device at the BS is $\gamma_{\mathrm{w}}$. With an adequate number of RBs allocated for the random weight transmission in the data transmission phase, we assume that the BS can successfully detect the weight sequences and determine the number of devices which have selected the same weight.
\begin{figure}[t]
\centering
\includegraphics[scale=0.30]{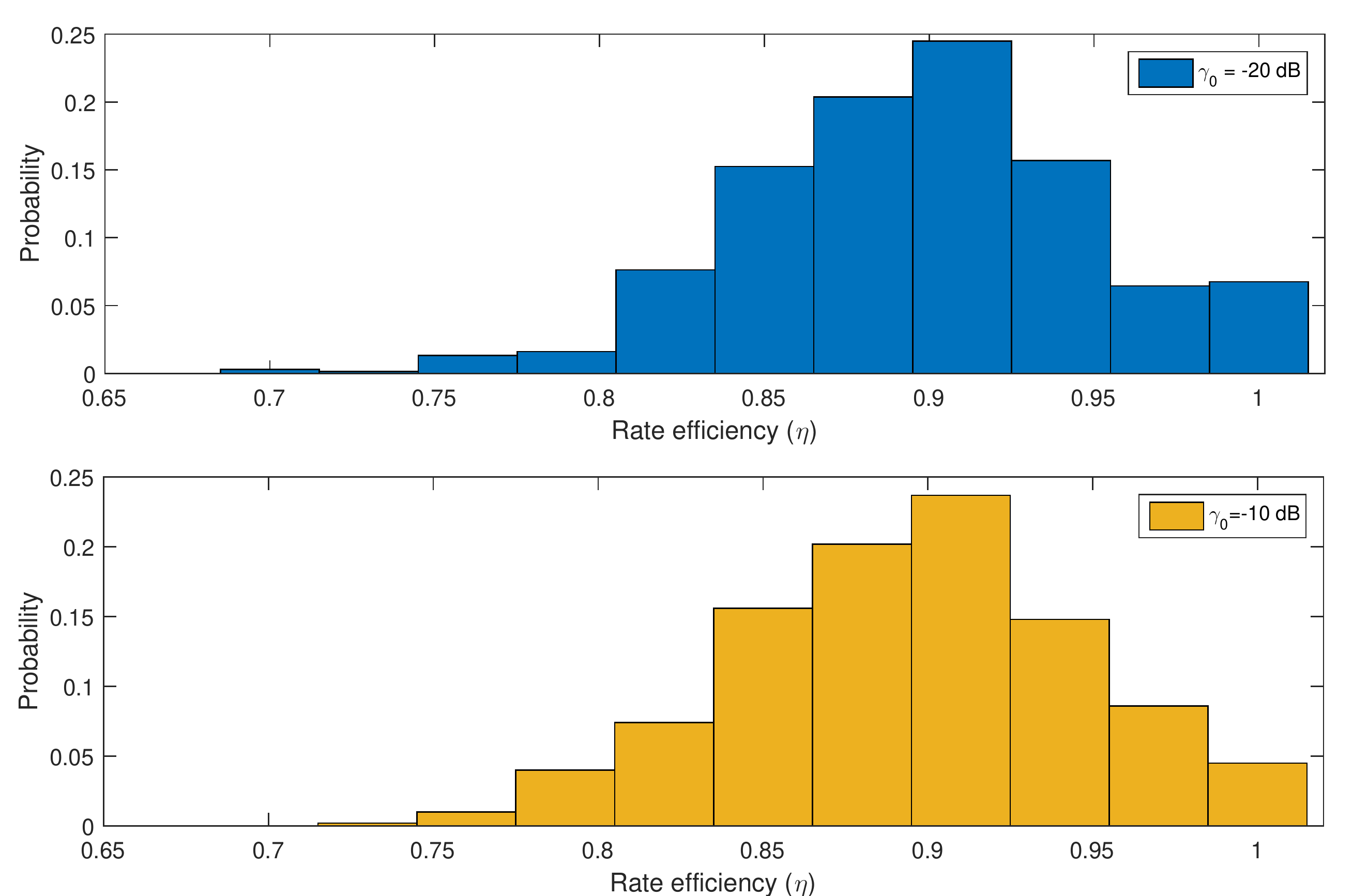}
\vskip-2ex
\caption{Histogram of the rate efficiency of a Raptor code with a message length of 1kb and optimized degree distribution in the low SNR regime.}
\label{histfigL1}
\end{figure}
\subsection{Rate Loss Due to the Small Packet Size of MTC Devices}
The capacity approaching degree distribution of Raptor codes in the low SNR regime was designed in \cite{LowSNRRaotor_Mahyar} for messages of infinite length, where the tree assumption of the bipartite graph of the Raptor code easily holds. However, one could use this degree distribution for finite message sizes with slight rate losses in the low SNR regime. Fig. \ref{histfigL1} shows the histogram of the rate efficiency of a Raptor code with the optimized degree distribution designed in \cite{LowSNRRaotor_Mahyar} for different SNR values, when the message size is  1024 bits. As can be seen in this figure, the rate efficiency of the finite message length Raptor code can be as small as 0.6 when the SNR is -10 dB. This however has a negative effect on the overall rate efficiency of the multi-layer Raptor code, as the overall rate efficiency is characterized by the minimum rate efficiency over all the layers. In the next section, we show that even with this imperfect rate efficiencies, the proposed scheme significantly outperforms the existing massive access strategies for M2M communications in terms of throughput and access delay. In fact the superiority of the proposed code over existing schemes comes from the simultaneous multiple device transmission over the same data channel, which significantly reduces the access delay and resource wastage.

\subsection{Adaptive Power Management at MTC Devices}
As can be seen in Fig. \ref{Sim1Fig}, the throughput per RB in the proposed scheme reduces when the number of active devices is small. This is because the designed target SNR has been set to a very small value, so several RBs are needed for an MTC device to successfully transmit its message at the very low rates required at very low SNRs. This leads to very low efficiencies if the arrival rate is low. To overcome this problem, the BS sends a power update message in the RAR message to inform the devices to update their transmission powers. This way when the number of active devices is low the devices transmit with higher powers such that each RB can support at least one MTC device. To minimize the number of RBs required in the data transmission phase, we adjust the target SNR based on the number of active devices such that the total received SNR at the BS is less than a threshold value, denoted by $\gamma_{\max}$. At the same time, and to limit the maximum transmission power of each device, we put a limit on the maximum allowable target SNR per MTC device. The target SNR, $\gamma_0$, can then be updated and informed to the devices by the BS as follows:
\begin{align}
\gamma_0=\min \{\sqrt[L]{1+\gamma_{\max}}-1, \gamma_{0,\max}\}.
\end{align}
The BS then needs to control the number of devices which are transmitting in the data transmission phase. For this aim, the BS estimates the maximum number of devices that can transmit at the same data channel for a given maximum allowable SNR at the BS, and then sends a not-to-transmit message to the devices which have selected some of the preambles. This way the BS can always effectively delay some of MTC devices for transmission at future time frames. In fact, conventional massive access management techniques can be combined with the proposed scheme to manage a large number of devices in M2M communications in an adaptive manner.
\section{Simulation Results}
For the simulations, we assume that the message size of each MTC device is $k=1024$ bits, including the device ID. Each device uses a Raptor code with a degree distribution function with maximum degree $300$ from \cite{LowSNRRaotor_Mahyar}, which has been optimized for Raptor codes in the low SNR regime and is given by $\Omega(x)=0.0174x+0.3488x^2+0.2309x^3+0.0695x^4+0.0873x^5+0.0002x^6+0.0805x^7+0.0004x^8+0.0191x^{11}+0.0518x^{12}+0.0123x^{23}+0.0310x^{24}+0.0220x^{59}+0.0020x^{60}+0.0268x^{300}$.  A rate 0.98 LDPC code is also used as the precoder for the Raptor code. We also assume that each RB has bandwidth $W_{\mathrm{s}}=1$ MHz and time duration $\tau_{\mathrm{s}}=1$ ms.

\begin{figure}[t]
\centering
\includegraphics[scale=0.28]{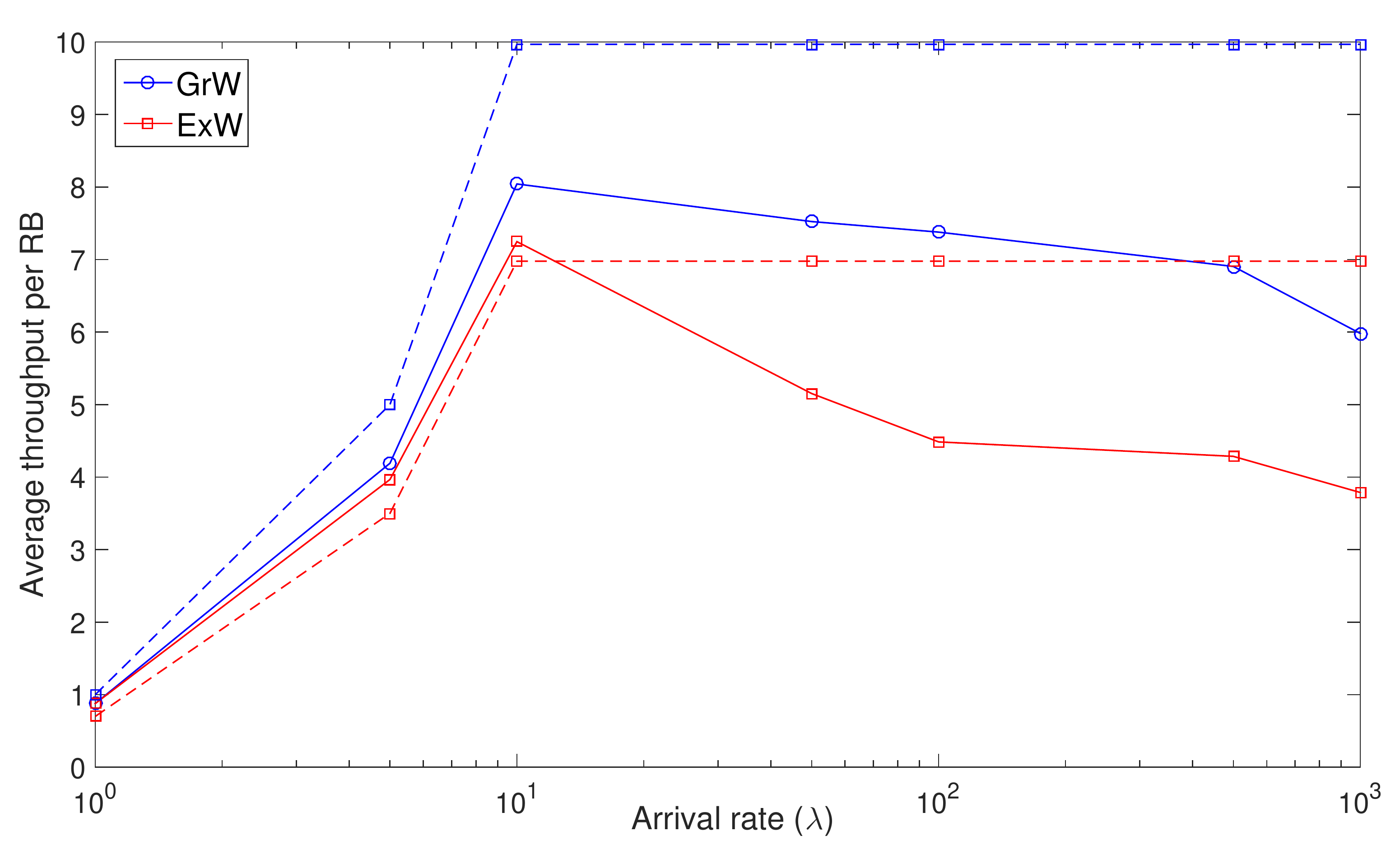}
\vskip-2ex
\caption{Average throughput per RBs for the proposed scheme with different weighting strategies with adaptive power management at the devices. The message length of MTC devices is $k=1024$, $N_\mathrm{s}=N_\mathrm{t}=20$, $W_\mathrm{s}=1$ MHz, and $\tau_{\mathrm{s}}=1$ ms. The maximum received SNR at the BS is $\gamma_{\max}=30$ dB. Solid and dashed lines show simulation and analytical results, respectively. }
\label{Sim2Fig}
\end{figure}

Fig. \ref{Sim2Fig} shows the average throughput per RB when a power adaptation strategy is used at the MTC devices such that the maximum received SNR at the BS is 30 dB. As can be seen in this figure in high load cases each RB can support up to 7 and 5 MTC devices per RB in GrW and ExW weighting strategies, respectively.

We compare the proposed scheme with the optimal access class barring (ACB) scheme \cite{ACB_Time}, which uses timing advance information in the random access phase to increase the MTC access probability. In this scheme, the BS can detect the timing advance information of multiple devices which have selected the same preamble. The BS then randomly selects one of these timing advances and includes it in the RAR message. If only one of the MTC devices, which have chosen the same preamble, has the selected timing advance, then it can send its message without collision.  This scheme can slightly improve the random access efficiency, but is still limited by the number of preambles, as the devices cannot simultaneously transmit at the same RB. In the ACB scheme, a parameter called ACB parameter, is broadcasted by the BS to inform the devices to transmit with a certain probability. Let $p$ denote the ACB parameter, then an MTC device which has data to transmit will draw a random number in the range $[0,1]$, and participate in the random access procedure only if the random number is less than $p$.  The optimal value for $p$ for the original ACB scheme \cite{DynamciACB} is $p=\min\{1,N/N_\mathrm{s}\}$ and that for the ACB with timing advance information \cite{ACB_Time} is $p=\left\{1,\frac{1.17N_\mathrm{s}\ln(\rho)}{N(\rho-1)}\right\}$, where $\rho=\frac{4d(R-d)}{R^2}$.
\begin{figure}[t]
\centering
\includegraphics[scale=0.28]{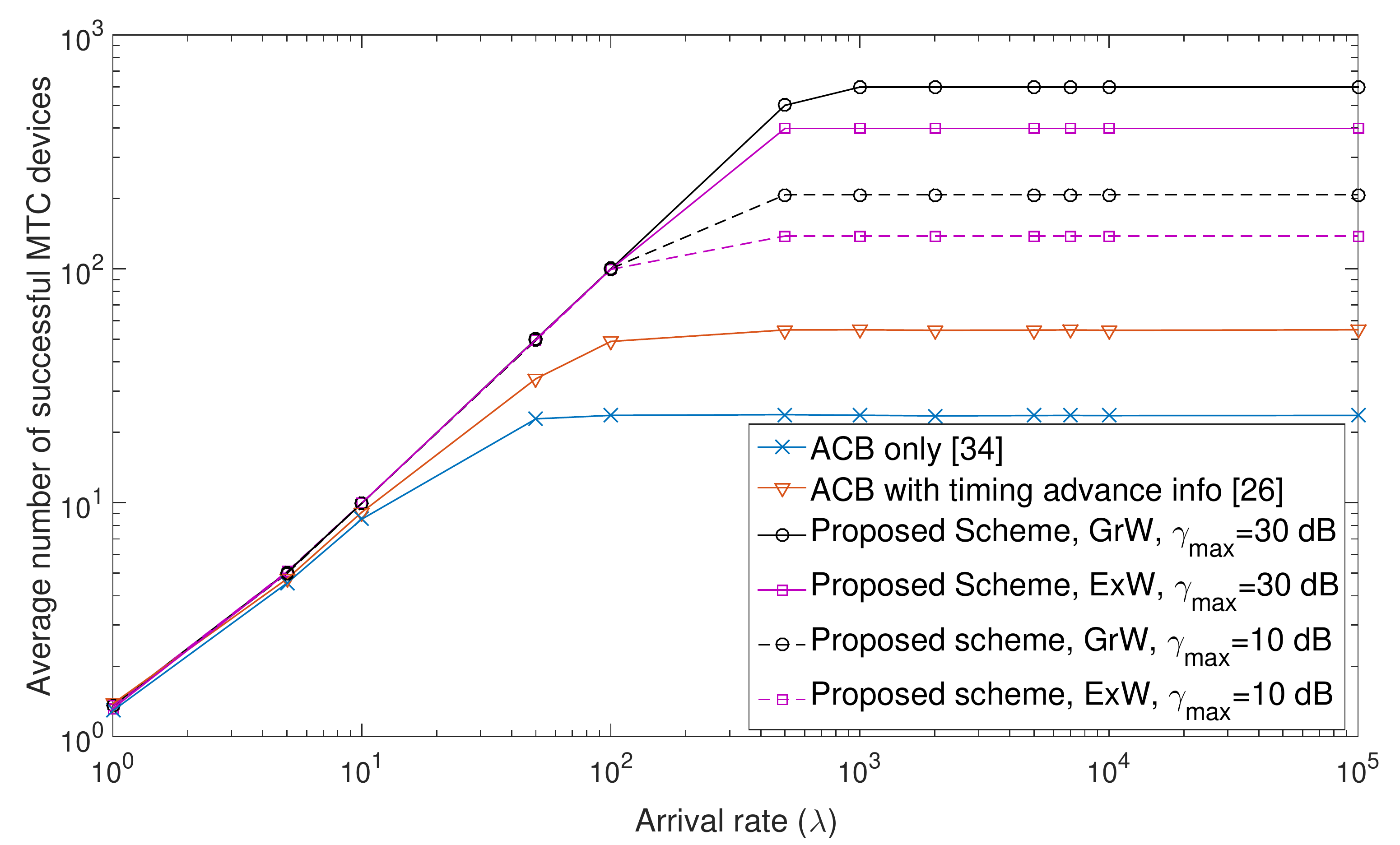}
\vskip-2ex
\caption{Average number of MTC devices that can be supported in each time frame for different massive access strategies. The message length of MTC devices is $k=1024$, $N_\mathrm{s}=64$, $N_\mathrm{t}=20$, $W_\mathrm{s}=1$ MHz, and $\tau_{\mathrm{s}}=1$ ms.}
\label{Sim3Fig}
\end{figure}

Similar to \cite{ACB_Time}, we assume that the cell has radius $R=1.5$ km, $\tau=0.26~\mu$s, and $d=c\tau=75$ m. This results in $N_\mathrm{s}=20$ different timing groups. Without loss of generality, we assume that each RB carries exactly $k_\mathrm{r}$ symbols, where $k_\mathrm{r}\ge k$, which means that in the ACB scheme, each device which has been granted access to the BS, will be allocated one RB. Fig. \ref{Sim3Fig} shows the average number of MTC devices which can be supported by different massive access techniques when 100 RBs are allocated for M2M communications in the data transmission phase. As can be seen in this figure the ACB scheme \cite{DynamciACB} can support at most 25 MTC devices while the ACB with timing advance information can support up to 55 MTC devices within each time frame. The proposed scheme significantly outperforms the ACB schemes by supporting up to 200 and 500 MTC devices when the maximum SNR at the BS is 10 dB and 30 dB, respectively.

\begin{figure}[t]
\centering
\includegraphics[width=8.5cm,height=5.5cm]{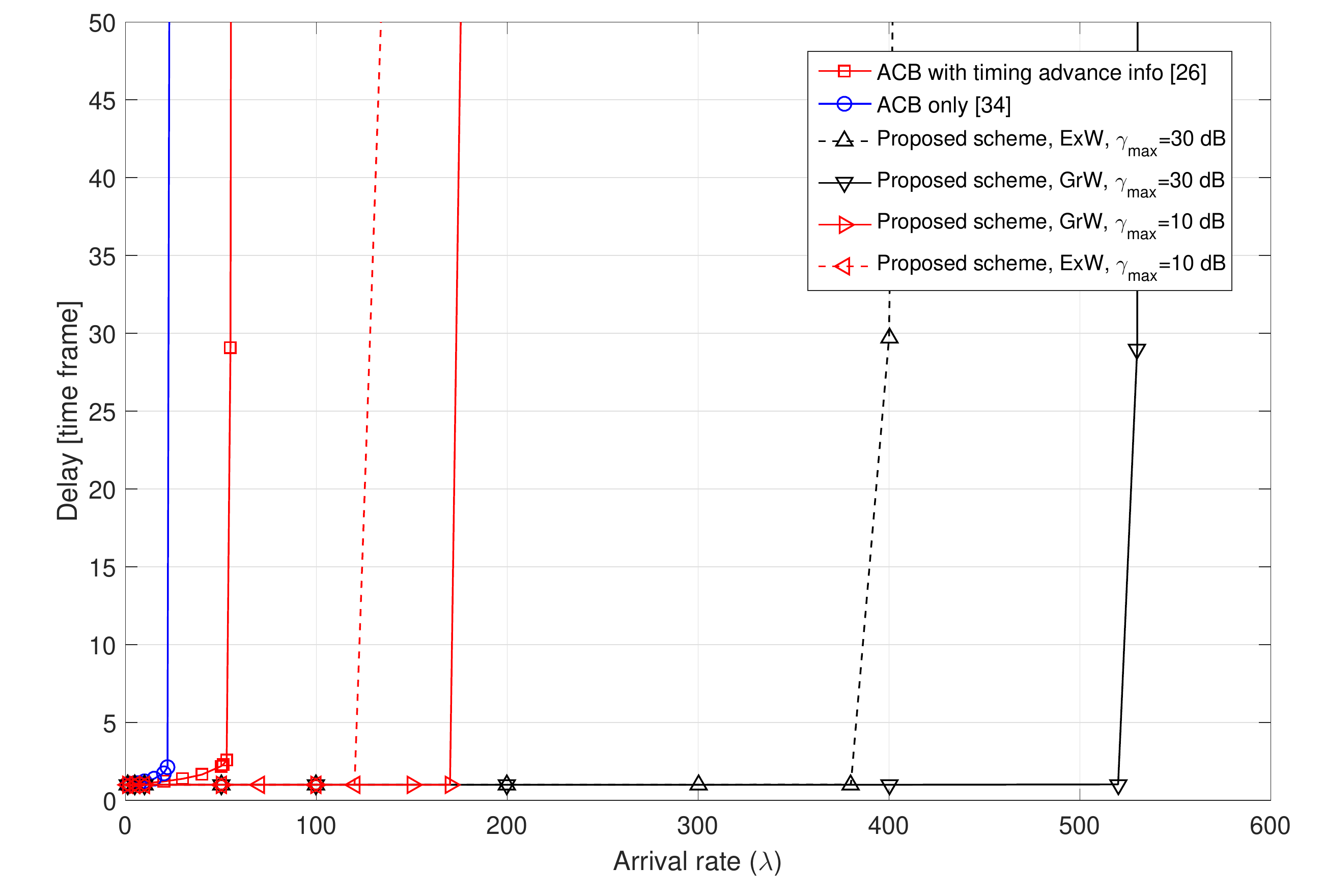}
\vskip-2ex
\caption{Average delay versus arrival rate. The message length of MTC devices is $k=1024$, $N_\mathrm{s}=64$, $N_\mathrm{t}=20$, $W_\mathrm{s}=1$ MHz, and $\tau_{\mathrm{s}}=1$ ms.}
\label{Sim4Fig}
\end{figure}

Fig. \ref{Sim4Fig} shows the average delay versus the arrival rate for different massive access strategies. As can be seen in this figure the proposed scheme can support a significantly larger number of devices with almost zero delay compared to ACB schemes. It is important to note that in the ACB scheme, we assume that a device can successfully deliver its message to the BS in its corresponding data channel, when it has successfully completed the RA phase; thus, the only limiting factor for the ACB scheme is the preamble collision in the RA phase. In ACB and to accommodate more devices in given number of RBs, each device should be allocated only few number of RBs and transmit at high power to guarantee that its message can be decoded. On the other hand in the proposed scheme, RBs are shared among all the devices and they are transmitting at minimum power over relatively large number of RBs. As the devices will only perform random access once in the proposed scheme and also according to Fig. \ref{Sim3Fig} where a larger number of MTC devices can be supported in each RB in the proposed scheme with lower transmission power, we can claim that the proposed scheme will significantly improve energy efficiency in M2M communications compared to existing ACB approaches.

\section{Conclusions}
In this paper, we designed an effective massive access strategy for highly dense cellular networks with M2M communications, where several MTC devices are able to simultaneously transmit in the same resource block by incorporating Raptor codes and superposition modulations. This significantly reduces the access delay and access blockage probability, which are the main burden in current proposals for random access management in M2M communications. A simple yet efficient random access strategy was proposed to only detect the selected preambles and the number of devices which have chosen them. The proposed scheme was analyzed and the maximum number of MTC devices that can be supported in a resource block was also characterized as a function of the message length, number of available resources, and the number of preambles. Simulation results showed that the proposed scheme can effectively support a massive number of M2M devices for a limited number of available resources, when the message size is small.

\bibliographystyle{IEEEtran}
\footnotesize
\bibliography{IEEEabrv,sample2}

\end{document}